\documentclass{mn2e}
\input{epsf}
\usepackage{amssymb}

\newbox\grsign \setbox\grsign=\hbox{$>$} \newdimen\grdimen \grdimen=\ht\grsign
\newbox\simlessbox \newbox\simgreatbox
\setbox\simgreatbox=\hbox{\raise.5ex\hbox{$>$}\llap
     {\lower.5ex\hbox{$\sim$}}}\ht1=\grdimen\dp1=0pt
\setbox\simlessbox=\hbox{\raise.5ex\hbox{$<$}\llap
     {\lower.5ex\hbox{$\sim$}}}\ht2=\grdimen\dp2=0pt

\newcommand{\hMpc}{{\ifmmode{h^{-1}{\rm Mpc}}\else{$h^{-1}$Mpc }\fi}}
\newcommand{\hkpc}{{\ifmmode{h^{-1}{\rm kpc}}\else{$h^{-1}$kpc }\fi}}
\newcommand{\hMsun}{{\ifmmode{h^{-1}{\rm {M_{\odot}}}}\else{$h^{-1}{\rm{M_{\odot}}}$}\fi}}
\newcommand{\Msun}{{\ifmmode{{\rm {M_{\odot}}}}\else{${\rm{M_{\odot}}}$}\fi}}

\title[The distribution function of dark matter]
{The distribution function of dark matter in massive haloes}
\author[R. Wojtak et al.]{Rados{\l}aw Wojtak,$^{1}$ Ewa L. {\L}okas,$^{1}$
Gary A. Mamon,$^{2,3}$ Stefan Gottl\"ober,$^{4}$\newauthor
Anatoly Klypin$^{5}$ and Yehuda Hoffman$^{6}$
\\   \\
$^1$Nicolaus Copernicus Astronomical Center, Bartycka 18, 00-716 Warsaw, Poland\\
$^2$Institut d'Astrophysique de Paris (UMR 7095: CNRS and Universit\'e Pierre \& Marie Curie),
    98 bis Bd Arago,
    F-75014 Paris, France \\
$^3$GEPI (UMR 8111: CNRS and Universit\'e Denis Diderot), Observatoire de Paris,
    F-92195 Meudon, France \\
$^4$Astrophysikalisches Institut Potsdam, An der Sternwarte 16, 14482 Potsdam, Germany\\
$^5$Department of Astronomy, New Mexico State University, Box 30001, Departament 4500,
Las Cruces, NM 880003, USA\\
$^6$Racah Institute of Physics, Hebrew University, Jerusalem 91904, Israel\\
}

\begin{document}

\maketitle

\begin{abstract}
We study the distribution function (DF) of dark matter particles in haloes
of mass range $10^{14}$--$10^{15}\Msun$. In the numerical part of this work
we measure the DF for a sample of relaxed haloes formed in the
simulation of a standard $\Lambda$CDM model. The DF is expressed as a
function of energy $E$ and the absolute value of the angular momentum $L$,
a form suitable for comparison with theoretical models.
By proper scaling we obtain the results that do not depend on the
virial mass of the haloes.
We demonstrate that the DF can be separated into
energy and angular momentum components and
propose a phenomenological model of the DF in the form
$f_{E}(E)[1+L^{2}/(2L_{0}^{2})]^{-\beta_{\infty}+\beta_{0}}L^{-2\beta_{0}}$.
This formulation
involves three parameters describing the anisotropy profile
in terms of its asymptotic values ($\beta_{0}$ and $\beta_{\infty}$)
and the scale of transition between them ($L_{0}$). The energy part
$f_{E}(E)$ is obtained via inversion of the integral for spatial density.
We provide a straightforward numerical scheme for this procedure as well as
a simple analytical approximation for a typical halo formed in the simulation.
The DF model is extensively compared with the simulations:
using the model parameters obtained from
fitting the anisotropy profile, we recover the DF from the simulation
as well as the profiles of the dispersion and kurtosis of radial and
tangential velocities. Finally, we show that our DF model reproduces
the power-law behaviour of phase space density
$Q=\rho(r)/\sigma^{3}(r)$.
\end{abstract}

\begin{keywords}
galaxies: clusters: general -- galaxies: kinematics and dynamics -- cosmology: dark matter
\end{keywords}

\section{Introduction}

The distribution function (DF) provides the most general and complete way of
statistical description of dark matter (DM) haloes. It carries maximum information
on the spatial and velocity distributions of particles in such objects.
Our knowledge on the DF is still being improved, mostly due
to numerical experiments. In the last few years cosmological
simulations have revealed increasingly detailed features of phase-space
structure of DM haloes. These numerical results provide useful
constraints on theoretical models of the DF. One property of interest
in this field is the anisotropy of the velocity
dispersion tensor. It has been demonstrated that the outer parts of the haloes
exhibit more radially anisotropic trajectories than the halo centre (see e.g.
Col\'in, Klypin \& Kravtsov 2000; Fukushige \& Makino 2001; Wojtak et al. 2005; Mamon \&
{\L}okas 2005; Cuesta et al. 2007). This feature, besides the well-studied
density profile, has been considered as the main point of reference in the
attempts at construction of a reliable model of the DF.

So far, a few
approaches to this problem have been proposed. Cuddeford (1991) generalized the
Osipkov-Merritt model (Osipkov 1979; Merritt 1985) to the DF
which generates an arbitrary anisotropy in the halo centre and becomes fully
radial at infinity. Although an analytical inversion for these models exists,
the anisotropy profile cannot be reconciled with the numerical results: 
the rise from central to outer anisotropy is too sharp and the outer 
orbits are too radial (see Mamon \& {\L}okas 2005).
An \& Evans (2006a) noticed that a non-trivial profile of
the anisotropy can be obtained from a sum of DFs
with a constant anisotropy for which an analytical inversion is known
(Cuddeford 1991; Kochanek 1996; Wilkinson \& Evans 1999). However, the resulting
anisotropy profiles are decreasing functions of radius and do not agree with those
measured in cosmological simulations. Recently a very elegant method
has been presented by Baes \& van Hese (2007). The authors introduced a
general ansatz for the anisotropy profile and then, for a given potential-density
pair, derived the DF as a series of some special functions.
This approach works well under the condition that the potential
can be expressed as an elementary function of the corresponding density.
This requirement, however, is not satisfied by many models, including the NFW
density profile (Navarro, Frenk \& White 1997) which is commonly used as a good
approximation of the universal density profile of DM haloes.

The DF inferred from the simulation gives a possibility to
test directly the analytical models. According to the Jeans theorem any
spherically symmetric system in the state of equilibrium should possess a DF
which is a function only of energy and the absolute value of angular
momentum. This theoretical postulate was taken into account in the computation
carried out by Voglis (1994) and Natarajan, Hjorth \& van Kampen (1997).
In the first case the DF was obtained for a single relaxed halo which formed
from cosmologically consistent initial conditions. It was shown that there were
two main contributions to the DF, the halo population and the core population of particles.
Both were effectively described by two independent phenomenological
fits. Natarajan et al. (1997) determined the DF for a sample of cluster-size haloes
formed in cosmological simulations. Their selection of objects included those with
substructures and departing from equilibrium.
They also discussed and took into account in their calculation the
effect of boundary conditions defined by the virial sphere. However, the final
results were not used to test quantitatively any model of the DF.

It seems that two main approaches to study the DF, namely
theoretical modelling and feedback from the simulations, evolved rather
separately barely crossing each other. The rare exceptions include the work of
{\L}okas \& Mamon (2001) who used the Eddington formula to derive numerically
the DF following from the NFW profile in the isotropic case and that of Widrow (2000)
who considered more general cuspy profiles and Osipkov-Merritt anisotropy.
This paper is devoted to combining both
approaches and providing a coherent analysis of the DF
from the viewpoints of the simulations as well as the model construction.
Our main aim is to propose a phenomenological model of the DF
that recovers the results from the simulations as accurately as possible.

Our effort is mainly motivated by the future applications of the derived DF to the
dynamical modelling of galaxy clusters. Although subhaloes in general have different
density and velocity distributions than DM particles (Diemand, Moore \& Stadel 2004),
massive subhaloes (those likely to host galaxies) are distributed like DM particles
(Faltenbacher \& Diemand 2006). Although the correspondence between the massive
subhaloes and galaxies in real clusters remains to be proven, our results should
at least in principle be applicable to
kinematic data sets for galaxy clusters. The traditional approach to do such
modelling was to reproduce the velocity dispersion profile of the galaxies by solving the
Jeans equation (see e.g. Katgert, Biviano \& Mazure 2004). It is
well known however that from the dispersion alone one cannot constrain all the
interesting parameters (such as the virial mass, the concentration of the NFW profile
and anisotropy) because of the density-anisotropy degeneracy. One can break this
degeneracy by using the fourth order velocity moment, the kurtosis, and solving an
additional higher-order Jeans equation ({\L}okas 2002; {\L}okas \& Mamon 2003;
{\L}okas et al. 2006; Wojtak \& {\L}okas 2007). Although this approach
has many advantages (e.g. it does not require the knowledge of the
full DF), it has been applied till now only for constant-anisotropy models and
the calculation of velocity moments requires the
binning of the data in which some information is lost. Since the number of galaxies with
measured redshifts per cluster is still rather low (of the order of a few hundred for the best-studied,
nearby clusters) it is essential that all the available information is used. This can be
obtained by fitting the projected DF to the data directly.

A few approaches along these lines have been attempted already. For example, Mahdavi \& Geller
(2004) used a simple DF of the form $f(E, L) \propto E^{\alpha-1/2} L^{-2 \beta}$
(which yields constant anisotropy) to
constrain the mass profile and orbital structure using combined kinematic data sets for
nearby galaxy groups and clusters, while van der Marel et al. (2000) in their study of
CNOC clusters did not assume an explicit
form for the energy-dependent part of the DF, but still used the constant anisotropy.
Wojtak et al. (2007) used a simplified form of
the projected isotropic DF constructed from the projected density combined with a Gaussian distribution
for the line-of-sight velocities to study the properties of members versus interlopers in
simulated kinematic data sets. None of the DFs used so far, however, reflects
accurately the true properties of cluster-size DM haloes found in $N$-body simulations.


The paper is organized as follows. Section 2 provides the theoretical framework and
defines all the basic quantities used later on in the paper. In the
next section we discuss the details of the computation of the DF
of DM particles in the haloes formed from cosmological simulations and provide examples
of the results. Section 4 is devoted to the derivation of
a phenomenological model of the DF; we discuss the separability of the DF in energy and angular
momentum and present an explicit formula for the $L$-dependent part of DF.
An extensive comparison of the model with the simulations
is presented in Section 5, where we also provide an analytical approximation for the
energy-dependent part of the DF obtained for an average halo.
Finally, the discussion follows in Section 6.

\vspace{-0.2in}

\section{Theoretical framework}

This section summarizes the theoretical background of the paper.
First, we introduce scaling properties consistent with the NFW density profile.
We will use this profile in the paper, but our approach is not restricted
to this particular density distribution and can be easily generalized to any
profile consistent with simulations (see below).
Second, we briefly describe the relation between the differential DF
and the DF itself. Finally, we discuss the consequences
of the finite volume of the virialized area of a halo. In particular, the relation
between the DF and its differential form is properly modified to account for this
effect.

\vspace{-0.1in}

\subsection{Scaling properties}

It is a well known fact that the density profiles of DM haloes formed in
cosmological simulations exhibit striking similarity. NFW showed that
most of them are well fitted within the virial sphere
by the universal two-parameter profile which can be expressed in the following
way
\begin{equation}\label{rho_NFW}
	\rho_{\rm NFW}(x)=\frac{1}{4\pi(\ln2-1/2)} \frac{M_{\rm s}}{r_{\rm s}^{3}}
	\frac{1}{x(1+x)^{2}},
\end{equation}
where $x=r/r_{\rm s}$. The two free parameters are the scale radius $r_{\rm s}$ and the
mass enclosed within the sphere of this radius $M_{\rm s}$. The (positive) gravitational
potential inferred from the Poisson equation reads
(Cole \& Lacey 1996; see also {\L}okas \& Mamon 2001)
\begin{equation}\label{psi_NFW}
	\Psi_{\rm NFW}(r)=V_{\rm s}^{2}\frac{\ln(1+x)}{x},
\end{equation}
where the velocity unit $V_{\rm s}$ is related to the circular velocity $V_{\rm cir}(r_{\rm s})$
at the scale radius via $V_{\rm s}=V_{\rm cir}(r_{\rm s})(\ln2-1/2)^{-1/2}$.

Let us note that $r_{\rm s}$, $V_{\rm s}$ and $M_{\rm s}$ define a set of natural units
of the NFW model. By scaling any quantity by a proper combination of them we
remove the explicit dependence on the free parameters of the NFW model.
This is an essential property if we want to study the dynamics
of a whole class of haloes with NFW-like density profiles.
Hereafter, we will keep this scaling in all equations in the text.
In many places we will also use a unit of the angular momentum
$L_{\rm s}$ as a substitute for $V_{\rm s}r_{\rm s}$.

\subsection{The distribution function}

The DF is a fundamental concept in statistical mechanics of
$N$-body systems. It describes the phase-space density of particles
of such a system
without any detailed knowledge of the time evolution of $N$ trajectories.
Following the Jeans theorem, a steady-state DF, which is of
interest for us here, depends on the phase-space coordinates only through the integrals of
motion. Although the shape of DM  haloes is in general better approximated by a three-axial ellipsoid
rather than a sphere
(see e.g. Gottl\"ober \& Yepes 2007), it is still very effective to assume spherical symmetry in dynamical approach.
Given that the streaming motions and internal rotation within the
virial sphere are negligible compared to higher velocity moments, spherical symmetry
implies that the DF can be expressed as
\begin{equation}\label{DF_sph}
	f({\mathbf r},{\mathbf v})=f(E,L),
\end{equation}
where $E$ is the positively defined binding energy and $L$ the
absolute value of the angular momentum per unit mass
\begin{eqnarray}
	E & = & \Psi(r)-\frac{1}{2}{\mathbf v}^{2}		\label{E_def} \\
	{\mathbf L} & = & {\mathbf r}\times{\mathbf v}.		\label{L_def}
\end{eqnarray}
The gravitational potential in equation (\ref{E_def}) is related to the DF
through the Poisson equation
\begin{equation}
	\nabla^{2} \Psi({\mathbf r})=-4\pi G\int f({\mathbf r},{\mathbf v})
	\textrm{d}^{3}{\mathbf v}.
\end{equation}

The most natural and straightforward probe of $f(E,L)$ in numerical experiments is
the so-called differential DF defined in the following way
\begin{equation}
	N(E,L)=\frac{\textrm{d}^{2}M}{\textrm{d}E\textrm{d}L}.
\end{equation}
One may intuitively interpret this function as mass density in energy-angular momentum
space. The DF itself can be simply derived dividing $N(E,L)$
by the volume $g(E,L)$ of the hypersurface of constant energy and angular momentum
embedded in the phase space
\begin{equation}\label{N(E,L)}
	f(E,L)=\frac{N(E,L)}{g(E,L)}.
\end{equation}
It is easy to show that the volume of this hypersurface reads (see Appendix A)
\begin{equation} \label{g(E,L)}
	g(E,L)=8\pi^{2}LT_{r}(E,L),
\end{equation}
where $T_{r}(E,L)$ is the radial period of an orbit given by the following integral
over radius from the pericentre to the apocentre
\begin{equation}	\label{T_r}
	T_{r}=2\int_{r_{\rm p}}^{r_{\rm a}}\frac{\textrm{d}r}
	{\sqrt{2[\Psi(r)-E-L^{2}/(2r^{2})]}}.
\end{equation}

\begin{figure}
\begin{center}
    \leavevmode
    \epsfxsize=8cm
    \epsfbox[75 75 580 800]{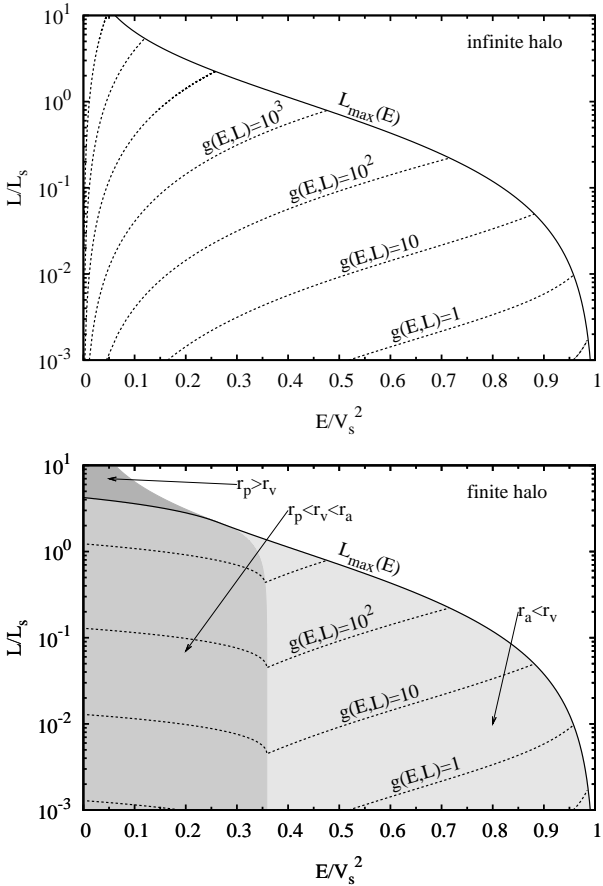}
\end{center}
\caption{The volume of the hypersurface of constant energy and angular momentum
for the infinite halo (upper panel) and the finite halo limited by the virial
sphere of radius $r_{\rm v}=5\,r_{\rm s}$ (lower panel). In both cases
the NFW profile was assumed. Solid lines show the profiles of the maximum angular
momentum of a given system. In the lower panel the three shades of gray mark the three
characteristic zones according to the orbit size defined by the relation of the virial
radius $r_{\rm v}$ to the pericentre radius $r_{\rm p}$
and the apocentre radius $r_{\rm a}$, as labelled.}
\label{g_vol}
\end{figure}

The upper panel of Fig.~\ref{g_vol} shows a contour map of $g(E,L)$
(dotted lines) calculated for the NFW gravitational potential
(\ref{psi_NFW}). The $L_{\rm max}(E)$ line is the profile
of maximum angular momentum which consists of points corresponding to circular
orbits. This curve divides the energy-angular momentum plane into an area
describing the physical orbits of a system (below $L_{\rm max}$) and the zone
not permitted by mechanics (above $L_{\rm max}$). Note that we
are using the scaling relations introduced in the previous subsection so
the results do not depend explicitly on the halo mass $M_{\rm s}$ and the scale radius
$r_{\rm s}$. In some places later on we will refer to the inverse function for
$L_{\rm max}(E)$ by $E_{\rm max}(L)$.

Voglis (1994) and Natarajan et al. (1997) showed that the dependence of $T_{r}(E,L)$
on the angular momentum is very weak and could be neglected without loss of
precision. This is understandable if we note that the NFW-like
potentials are still not so far away from the isochrone potential
$\Psi(r) \propto (b+\sqrt{b^{2}+r^{2}})^{-1}$ which leads to purely energy-dependent
$T_{r}$ proportional to $E^{-3/2}$ (Binney \& Tremaine 1987). Following Natarajan et al.
(1997) we will use this feature to simplify expression (\ref{g(E,L)}). To do this we
first note that the volume of the hypersurface of constant energy $g_{E}$ is given by
\begin{equation}\label{g_E}
	g_{E}(E)=\int_{0}^{L_{\rm max}(E)}g(E,L)\textrm{d}L.
\end{equation}
Taking advantage of the weak dependence of $T_r$ on $L$, equation (\ref{T_r}),
we get
\begin{equation}\label{g_app}
	g(E,L)\approx 2\,\frac{g_{E}(E)\,L}{L_{\rm max}^{2}(E)}.
\end{equation}
On the other hand, one can show that $g_{E}(E)$ reads (see Appendix A)
\begin{equation}\label{g_E_gen}
	g_{E}(E)=16\pi^{2}\int_{0}^{r_{\rm max}(E)}
	\sqrt{2(\Psi(r)-E)}r^{2}\textrm{d}r,
\end{equation}
where $r_{\rm max}(E)$ is the apocentre radius of the radial orbit. Inserting
(\ref{g_E_gen}) into (\ref{g_app}) one immediately gets a very simple
approximation for $g(E,L)$ involving only a one-dimensional integral without
singularities, in contrast with $g(E, L)$ derived by expression (\ref{T_r}).
We find that this
approximation reproduces the exact formula (\ref{g(E,L)}) with enough
accuracy. Taking advantage of its numerical simplicity we use it in majority
of our calculations.

\subsection{Boundaries of the haloes}

So far we have discussed the relation between the DF and
its differential form for an infinite system. In practice, however, we restrict
our numerical analysis to the interior of the virial sphere which
separates the equilibrium part of a halo from the infall region. We define
the virial radius $r_{\rm v}$ of this sphere by
\begin{equation}\label{r_v_m_v}
	M_{\rm v}=\frac{4}{3}\pi r_{\rm v}^{3}\Delta_{\rm c}\rho_{\rm c},
\end{equation}
where $M_{\rm v}$ is the virial mass, $\rho_{\rm c}$ is the present critical density
and $\Delta_{\rm c}$ is the virial overdensity. Another parameter commonly
used to describe the size of the virial sphere in terms of $r_{\rm s}$ is the
concentration $c=r_{\rm v}/r_{\rm s}$.

The existence of the boundary of the virialized part of the halo implies that $T_{r}$ given by
(\ref{T_r}) must be replaced by
\begin{equation}	\label{T_r_trun}
	T_{r}=2\int_{r_{\rm p}}^{{\rm min}\{r_{\rm a},r_{\rm v}\}}\frac{\textrm{d}r}
	{\sqrt{2[\Psi(r)-E-L^{2}/(2r^{2})]}},
\end{equation}
where the upper limit of the integral is a minimum of the virial radius and
the radius at the apocentre (see Appendix A for details). Combining (\ref{T_r_trun})
with (\ref{g(E,L)}) one gets a general formula for the volume $g(E,L)$ in the
presence of a spherical boundary of a halo. Contrary to the conclusion of
Natarajan et al. (1997), we find that the approximation (\ref{g_app}) is no longer
justified for orbits extending beyond the virial sphere ($r_{\rm a}>r_{\rm v}$).
This follows from the fact that angular momentum dependence of (\ref{T_r_trun})
becomes non-negligible and the integral (\ref{g_E}) cannot be simplified to the form
of (\ref{g_app}).

Using (\ref{g(E,L)}) and (\ref{T_r_trun}) with the NFW potential, we calculated
$g(E,L)$ for a halo limited by the virial sphere of radius $r_{\rm v}=5\,r_{\rm s}$
(see the lower panel of Fig.~\ref{g_vol}). As expected, the result differs from an
infinite system by the orbits with $r_{\rm a}>r_{\rm v}$ and remains unchanged for
trajectories wholly included within the virial sphere.

\section{The distribution function from the simulation}

\subsection{The simulation}

For our $N$-body simulation we have assumed the WMAP3 cosmology (Spergel et al. 2007)
with matter density $\Omega_{\rm m} = 0.24$, the cosmological constant
$\Omega_{\Lambda} = 0.76$, the dimensionless Hubble parameter $h = 0.73$,
the spectral index of primordial density perturbations $n = 0.96$ and the
normalization of power spectrum $\sigma_{8} = 0.76$. We have used a box of size $160 \: \hMpc$ and
$1024^3$ particles. Thus the particle mass was $3.5 \times 10^8 \Msun$.
Starting at redshift $z=30$ we followed the evolution using the MPI version of
the Adaptive Refinement Tree (ART) code (Kravtsov, Klypin \& Khokhlov 1997).

We identified clusters with the hierarchical friends-of-friends (FOF) algorithm (Klypin
et al. 1999) with a linking length of 0.17 times the mean inter-particle distance which
roughly corresponds to an overdensity of 330. We have selected 36 clusters at redshift
$z=0$ in the range of virial mass (0.15-2) $\times 10^{15} \Msun$, where the
virial overdensity parameter appropriate for our cosmological model
was assumed $\Delta_{\rm c}=93.8$ ({\L}okas \& Hoffman 2001).
Our sample did not include clusters with two substructures of approximately the same
mass and a poor fit of the NFW profile suggestive of a recent major merger.

Starting from the FOF position of the cluster we have determined the highest density
peak as the final centre of the clusters. This centre coincides with the position of the
most massive substructure found at the linking length 8 times shorter and also with the
position of the halo found by the BDM halo finder (Klypin et al. 1999).

\subsection{Computation of the distribution function}

In the first step of the computation we calculate the binding energy (\ref{E_def})
and the angular momentum (\ref{L_def}) per unit mass for each particle within the
virial sphere of each halo. Spherical symmetry implies that we have to apply
in (\ref{E_def}) the radial profile of the gravitational potential
\begin{equation}\label{pot_num_0}
	\Psi(r)=\int_{r}^{\infty}\frac{GM(r){\rm d}r}{r^{2}},
\end{equation}
where $\Psi(\infty)=0$. However, the mass profile of the equilibrium part of a halo
reaches no further than the virial radius. On the other hand, all analytical models
of the DF involve the density profile extending to infinity. We found that
the only coherent way to reconcile both facts is to split the integral
(\ref{pot_num_0}) in two parts
\begin{equation}\label{pot_num_1}
	\Psi(r)=\int_{r}^{r_{\rm v}}\frac{GM(r){\rm d}r}{r^{2}}+
	\int_{r_{\rm v}}^{\infty}\frac{GM_{\rm NFW}(r){\rm d}r}{r^{2}}.
\end{equation}
The first term is evaluated numerically by integration of a discrete mass
profile. The second term is an analytical extension with the NFW density
profile which is an assumption of the DF model introduced in the
following section. Its contribution to the potential is a constant equal to
$V_{\rm s}^{2}\ln(1+c)/c$.

\begin{figure}
\begin{center}
    \leavevmode
    \epsfxsize=8cm
    \epsfbox[75 75 580 420]{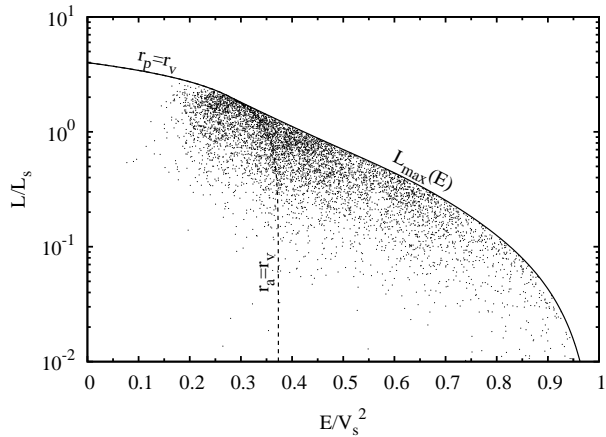}
\end{center}
\caption{Energies and angular momenta of particles within the virial sphere of one
of the simulated haloes. Solid and dashed lines mark the profile of the maximum angular
momentum and the limit of vanishing radial velocity at $r_{\rm v}$ respectively.
To make the picture less obscure we plotted only 1 percent of the particles.}
\label{halo_1_sc}
\end{figure}

Fig.~\ref{halo_1_sc} shows the resulting energies and angular momenta of
particles inside the virial sphere of one of the simulated haloes.
The profile of the maximum angular momentum (solid line) and the profile
of vanishing radial velocity at the virial sphere (dashed line) were calculated
for the exact gravitational potential given by (\ref{pot_num_1}). All
particles occupy the area permitted by mechanics or lie
very close to the boundary line. Interestingly, quite a large fraction of
them have orbits extending beyond the virial sphere.
As noted in the previous section, we keep $V_{\rm s}^{2}$ and $L_{\rm s}$ as units
of energy and angular momentum respectively. The parameters of the NFW
model were obtained for each halo by fitting the NFW formula to the density
profile measured in logarithmic radial bins.

In the next step we determine for each halo the differential DF
given by (\ref{N(E,L)}). In this calculation we used our own version of the FiEstAS
(Field Estimator for Arbitrary Spaces) algorithm designed to
infer the density field from a scatter diagram embedded in a space of any number of
dimensions (see Ascasibar \& Binney 2005 for more details). As a result of this
computation we get an estimate of $N(E,L)$ at all points of the energy-angular
momentum plane corresponding to the particles inside the virial sphere.
Once $N(E,L)$ is calculated the DF can be easily obtained
via (\ref{N(E,L)}). As discussed in section 2, we used approximation
(\ref{g_app}) for the orbits contained inside the virial sphere
and the exact formula (\ref{g(E,L)}) with (\ref{T_r_trun})
for trajectories extending beyond  $r_{\rm v}$. We found that the additional advantage
of expression (\ref{g_app}) is that it could be evaluated at any point of the energy-angular
momentum plane. This helps us to keep the estimates of the DF obtained for points
with angular momentum lying slightly above $L_{\rm max}(E)$.

\begin{figure}
\begin{center}
    \leavevmode
    \epsfxsize=8cm
    \epsfbox[75 75 580 800]{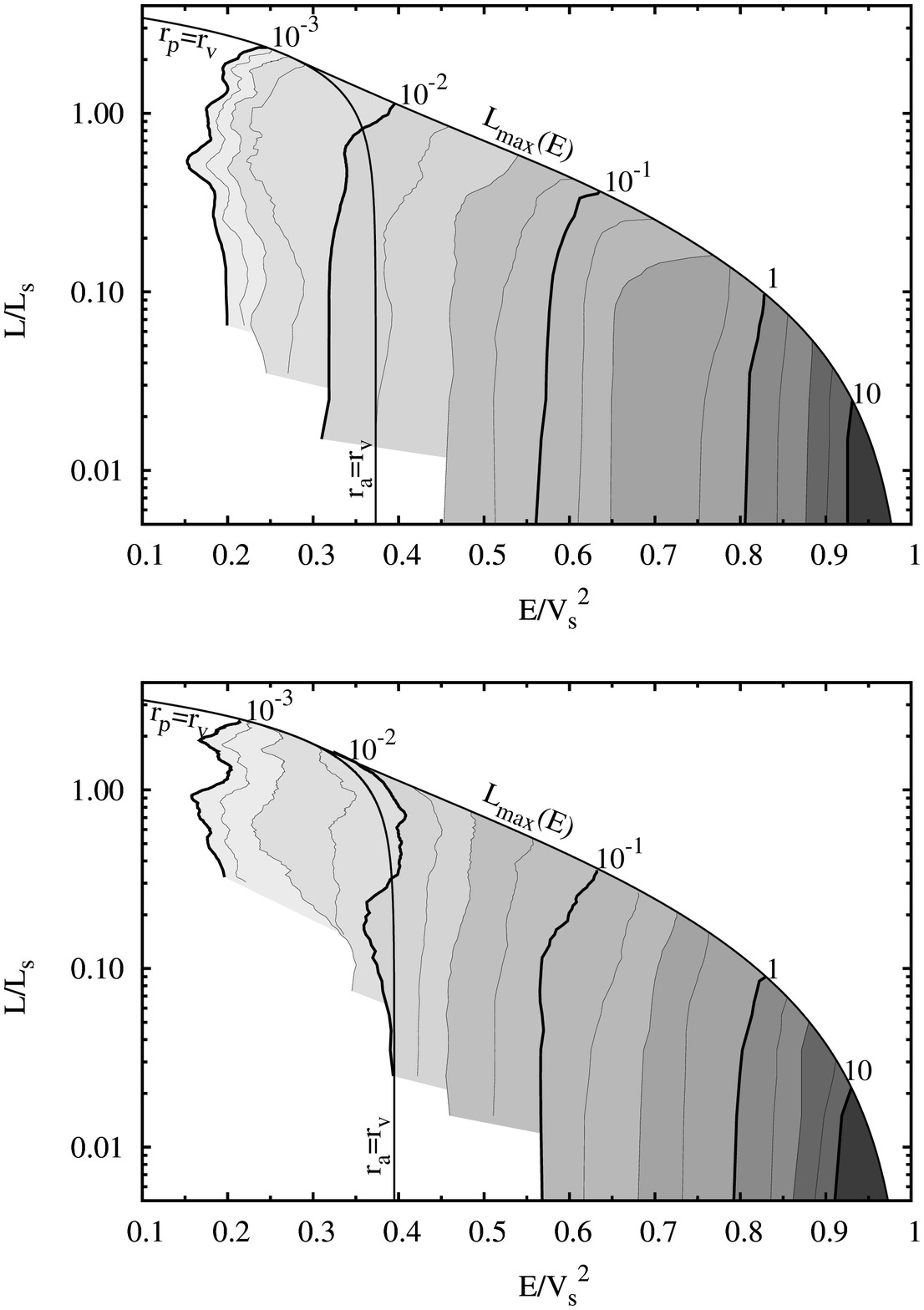}
\end{center}
\caption{Contour maps of the DF of DM particles inside
the virial sphere of two example haloes. The profile of the maximum angular momentum is
indicated by $L_{\rm max}(E)$ and the line of vanishing radial velocity at the virial
sphere by $r_{\rm a}=r_{\rm v}$ (apocentre at the virial sphere) or $r_{\rm p}=r_{\rm v}$ (pericentre
at the virial sphere).}
\label{halo_1_DF}
\end{figure}

In order to derive a contour map or a profile of the DF we introduce
a regular dense mesh on the energy-angular momentum plane and find the median value of
the DF in each cell. Such a set of median points is considered as the final numerical
approximation of the DF and is used in preparation of all plots in this paper.
Fig.~\ref{halo_1_DF} shows two examples of the resulting contour maps obtained for two
different haloes. The unit of the DF in this and following Figures
is $M_{\rm s}/r_{\rm s}^{3}/V_{\rm s}^{3}$.
The interval between the iso-DF lines is fixed at value
$0.25$ of the logarithmic scale. The lack of the DF estimation in the lower part of
each diagram arises from the fact that this zone is occupied by very few particles
(see e.g. Fig.~\ref{halo_1_sc}) so that no information on the distribution can be
retrieved. Let us note that this is an effect of the finite mass resolution.

\begin{figure}
\begin{center}
    \leavevmode
    \epsfxsize=8cm
    \epsfbox[75 75 580 420]{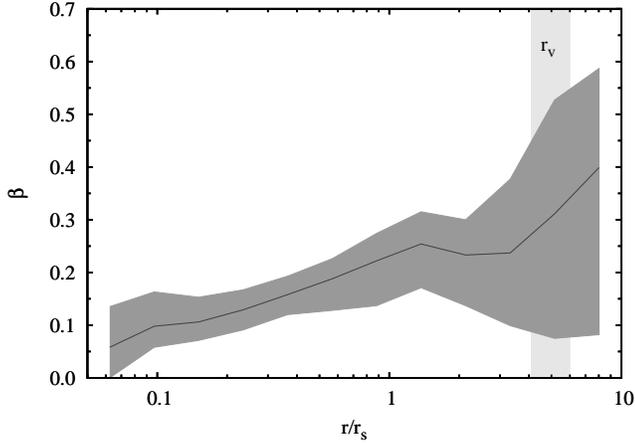}
\end{center}
\caption{The profile of the anisotropy parameter. The solid line and
the dark gray area are the median and the interquartile range of the profiles
obtained for individual haloes and rescaled by $r_{\rm s}$ inferred from
fitting the NFW profile.}
\label{beta_sing}
\end{figure}

\section{The analytical model of the distribution function}

A general form of the DF for spherical systems
in the state of equilibrium is a function of energy and the absolute
value of angular momentum $f(E,L)$. In our approach we assume that
the DF is separable in energy and angular momentum
\begin{equation}\label{DF_mod_gen}
	f(E,L)=f_{E}(E)f_{L}(L).
\end{equation}
This is the first assumption that considerably narrows the family
of possible solutions. Therefore, it is necessary to check how robust
it is. We address this problem in the next section, where we present
an extensive comparison of the analytical model with the simulations.

The angular momentum part of the DF in equation (\ref{DF_mod_gen})
specifies the anisotropy of velocity dispersion tensor. This quantity
is commonly described with the so-called anisotropy parameter
\begin{equation}\label{beta_def}
	\beta(r)=1-\frac{\sigma_{\theta}^{2}(r)}{\sigma_{r}^{2}(r)},
\end{equation}
where $\sigma_{r}$ and $\sigma_{\theta}$ are the radial and the
tangential velocity dispersions respectively and we assume there are no
streaming motions.
The values of this parameter range from $-\infty$ for circular orbits
to $1$ for purely radial trajectories. Fig.~\ref{beta_sing} shows
the average anisotropy profile of the simulated haloes used for the measurement
of the DF. The light gray rectangle in the background of the plot indicates the position of
the virial radius. It is clearly seen that the anisotropy is typically
a growing function of radius, with values $\sim 0.07$ in the halo centre
and $\sim 0.3$ at the virial sphere (see e.g. Mamon \& {\L}okas 2005
and Cuesta et al. 2007 for comparison). On the other hand,
the considerable width of the interquartile range of the measured $\beta(r)$
(dark gray region)
signifies that the profiles of single haloes differ among each other.
Occasionally flat or decreasing profiles are measured.
It seems that a simple and general enough analytical model of the anisotropy
should possess at least three free parameters which determine asymptotic
values of $\beta(r)$ for small and large radii and a scale
of transition between them. We proceed with the construction
of such a model by introducing a proper ansatz for $f_{L}(L)$.

Louis (1993) showed that the following asymptotes of the angular
momentum part of the DF
\begin{equation}\label{f_L_louis}
	f_{L}(L) \propto \left\{
	\begin{array}{ll}
	1 & \textrm{for $L\ll L_{0}$} \\
	L^{-2\beta_{\infty}} & \textrm{for $L\gg L_{0}$},
\end{array} \right.
\end{equation}
where $L_{0}$ is an angular momentum constant, lead to constant
anisotropy $\beta_{\infty}$ at infinity ($r^{2}\Psi(r)\gg L_{0}^{2}$)
and $\beta=0$ in the halo centre. This result can be easily generalized
to the case of a non-isotropic velocity distribution in both limits of
radius. First, let us note that the central part of the halo
is dominated by the particles with small angular momenta,
namely $L^{2}\le 2r^{2}\Psi(r)\ll L_{0}^{2}$.
Then, remembering that the DF of constant anisotropy takes the form
(H\'enon 1973; Binney \& Tremaine 1987; {\L}okas 2002)
\begin{equation}	\label{DF_const_beta}
	f(E,L)=f_{E}(E)L^{-2\beta},
\end{equation}
it is easy to notice that the formula (\ref{f_L_louis}) can be rewritten
in the following way
\begin{equation}	\label{f_L_asympt}
	f_{L}(L) \propto \left\{
	\begin{array}{ll}
	L^{-2\beta_{0}} & \textrm{for $L\ll L_{0}$} \\
	L^{-2\beta_{\infty}} & \textrm{for $L\gg L_{0}$},
\end{array} \right.
\end{equation}
where $\beta_{0}$ is the central anisotropy of a system. As shown by
An \& Evans (2006b), the upper limit for $\beta_{0}$ is equal to $\gamma/2$,
where $r^{-\gamma}$ is the density profile near the halo centre. This means
that for the NFW density model we have $\beta_{0}\le 1/2$.

The simplest function obeying the asymptotic conditions formulated above is a double
power-law function
\begin{equation}\label{f_L}
	f_{L}(L)=\Big(1+\frac{L^{2}}{2L_{0}^{2}}\Big)^{-\beta_{\infty}+\beta_{0}}
	L^{-2\beta_{0}}.
\end{equation}
As shown in the following section, this ansatz leads to a
very realistic anisotropy profile that fits well the $\beta(r)$
profiles of simulated haloes. Furthermore, the simplicity of formula
(\ref{f_L}) guarantees that the energy part of the DF
can be quite easily calculated via the inversion of the integral
equation
\begin{equation}\label{rho_DF}
	\rho(r)=\int\!\!\!\int\!\!\!\int f_{E}(E)
	\Big(1+\frac{L^{2}}{2L_{0}^{2}}\Big)^{-\beta_{\infty}+\beta_{0}}
	L^{-2\beta_{0}}\textrm{d}^{3}v.
\end{equation}
The key idea of this procedure lies in an analytical simplification of the
right-hand side of (\ref{rho_DF}) to a one-dimensional integral. The resulting
equation is then solved numerically for $f_{E}(E)$. The technical
details of this calculation are summarized in Appendix B. 
Once the full form of the DF is determined one can also calculate the
velocity moments. All formulae are reduced to one-dimensional integrals
which can be easily evaluated numerically (see Appendix C).

\begin{figure*}
\begin{center}
    \leavevmode
    \epsfxsize=17.5cm
    \epsfbox[70 50 860 420]{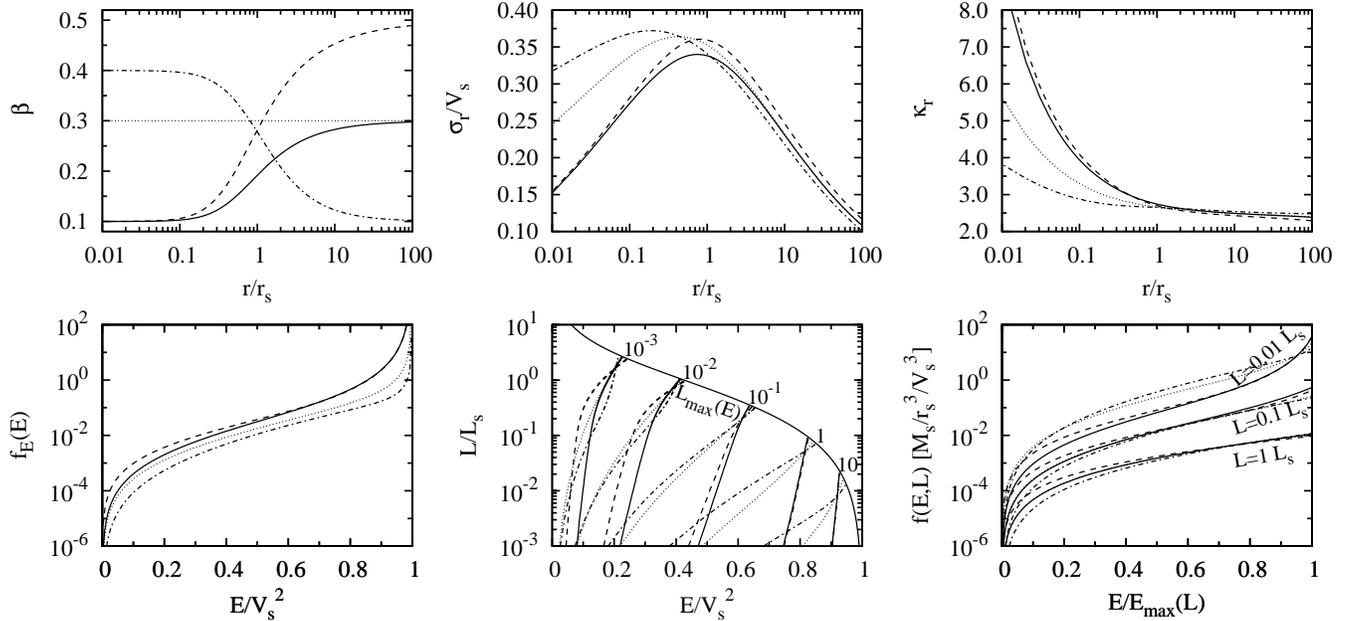}
\end{center}
\caption{The top panels show (from left to right) the anisotropy, dispersion and kurtosis
of the radial velocity inferred from the model of the DF
with four sets of parameters: $\beta_{0}=0.1$, $L_{0}=0.25\,L_{\rm s}$
and $\beta_{\infty}=0.3,\,0.5$ (solid and dashed lines respectively);
$\beta_{0}=\beta_{\infty}=0.3$, $L_{0}=0.25\,L_{\rm s}$ (dotted line); $\beta_{0}=0.4$,
$L_{0}=0.25\,L_{\rm s}$ and $\beta_{\infty}=0.1$ (dashed-dotted line). The corresponding
DFs for the same sets of parameters are plotted in the bottom panels in terms of:
the energy part of the DF $f_{E}(E)$ (left), iso-DF
lines with values indicated along the curve of maximum angular momentum
(middle) and the profiles of the DF for three values of angular momentum
(right). In all calculations the NFW density profile was assumed.}
\label{DF_th}
\end{figure*}

The top row of Fig.~\ref{DF_th} shows the anisotropy, dispersion $\sigma_{r}$
and kurtosis $\kappa_{r}=\langle v_{r}^{4}\rangle/\sigma_{r}^{4}$ of the
radial velocity inferred from the model of the DF. The
calculations were carried out assuming the NFW density profile and four
sets of model parameters chosen to illustrate the flexibility of the model:
$\beta_{0}=0.1$ and $\beta_{\infty}=0.3,\,0.5$
(solid and dashed lines respectively); $\beta_{0}=\beta_{\infty}=0.3$
(dotted line); $\beta_{0}=0.4$ and $\beta_{\infty}=0.1$ (dashed-dotted line).
In all cases the transition value of $L_{0}=0.25\,L_{\rm s}$ was used.

The dispersion profiles for the two models with increasing $\beta(r)$, as
expected, differ only for large radii which is the effect
of different values of $\beta_{\infty}$. Interestingly, the corresponding
kurtosis profiles clearly signify flat-topped velocity distribution
in the outer part of the halo ($\kappa_{r}<3$), highly peaked distribution in the
centre ($\kappa_{r}>3$) and roughly Gaussian for radii around $r_{\rm s}$
($\kappa_{r}\approx 3$). On the other hand, non-increasing $\beta(r)$ profiles
lead to less peaked velocity distributions in the centre.
It seems therefore that the typical anisotropy of DM haloes,
as shown in Fig.~\ref{beta_sing}, is expected to coincide with the
kurtosis rapidly growing towards the halo centre
(see also Fig.~\ref{4haloes} below). As we will see in the
following section, this is one of the most characteristic features
of the phase-space structure of massive DM haloes.

In the bottom panels of Fig.~\ref{DF_th} we plotted the DFs
corresponding to four sets of model parameters, as described above. The three panels
from the left to the right show the energy part of the DF, contour maps
and the profiles for three fixed values of angular momentum.
The plots reveal some interesting signatures of
the specific shape of $\beta(r)$ profile. For example,
the inclination of the iso-DF lines with respect to the energy axis
decreases with increasing $\beta_{0}$: more isotropic $\beta$ at the
centre corresponds to more vertical iso-DF lines; also the shape of the
lines is somewhat different.
These features are also to some extent visible in the contour maps
of the DF for two simulated haloes in Fig.~\ref{halo_1_DF}. The
upper map represents a halo with an increasing $\beta(r)$, whereas the
second one depicts the case of a decreasing $\beta(r)$ profile.
Both haloes are analyzed in terms
of velocity moments and the DF in the following section.

\begin{figure}
\begin{center}
    \leavevmode
    \epsfxsize=8cm
    \epsfbox[75 75 580 420]{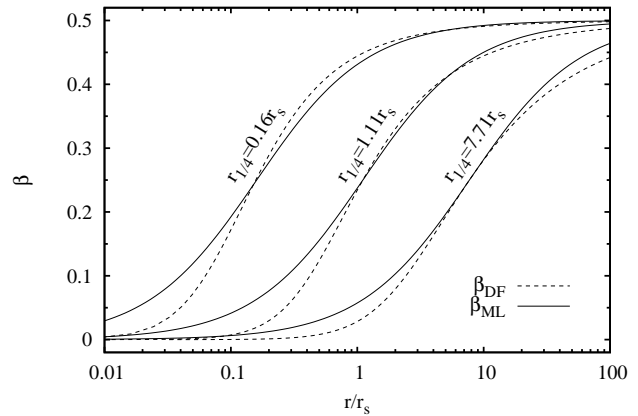}
\end{center}
\caption{Comparison between the functional form of the anisotropy
(\ref{ML_beta}) proposed by Mamon \& {\L}okas (2005) ($\beta_{\rm ML}$) and
$\beta(r)$ inferred from the DF model ($\beta_{\rm DF}$) with $\beta_{0}=0$
and $\beta_{\infty}=0.5$.}
\label{ML_beta_f}
\end{figure}

Recently Mamon \& {\L}okas (2005) found that the simulation data are well reproduced
by the anisotropy profile of the form
\begin{equation}	\label{ML_beta}
	\beta(r)=\frac{1}{2}\frac{r}{r+r_{1/4}},
\end{equation}
where $r_{1/4}$ is the radius where $\beta=0.25$.
Assuming $\beta_{0}=0$ and $\beta_{\infty}=0.5$ in our DF model, we made
a comparison of the resulting anisotropy with the functional form (\ref{ML_beta}).
Fig.~\ref{ML_beta_f} shows both anisotropies for three values of $r_{1/4}$.
Note that both $\beta(r)$ profiles have similar shapes, although our anisotropy 
profile has a somewhat sharper rise at small radii.

We also find
that the radius $r_{0}$ characteristic of the DF model, for which $\beta$
is the mean of the limiting values
\begin{equation}
	\beta(r_{0})=\frac{\beta_{0}+\beta_{\infty}}{2},
\end{equation}
depends weakly on $\beta_{0}$ and $\beta_{\infty}$. For parameter ranges
leading to $\beta(r)$ profiles covering the interquartile area of
anisotropy from the simulation ($0<\beta_{0}<0.15$, $0<\beta_{\infty}<0.6$
and $0.04<L_{0}/L_{\rm s}<25$), this radius is well (within 5 percent accuracy)
approximated by
\begin{equation}
	r_{0}/r_{\rm s}=3.69(L_{0}/L_{\rm s})^{0.97}+2.27(L_{0}/L_{\rm s})^{1.9}.
\end{equation}

\section{Comparison with the simulation}
\subsection{The distribution function}

The DF proposed in the previous section is a phenomenological
model in the sense that it possesses free parameters whose values should be adjusted
to the simulation data. All three parameters were introduced to determine a family
of anisotropy profiles so that it is $\beta(r)$ that is most sensitive to
the variations of $\beta_{0}$, $\beta_{\infty}$ and $L_{0}$. Consequently,
we decided to constrain the parameters of the model by fitting the $\beta(r)$ profile inferred
from the DF model to the median profile measured in simulated DM haloes. The best-fitting
parameters are: $\beta_{0}=0.09$, $\beta_{\infty}=0.34$ and $L_{0}=0.198\,L_{\rm s}$.
The corresponding best-fitting profile of the anisotropy is plotted as a dashed line
in the lower left panel of Fig.~\ref{moments}.

\begin{figure}
\begin{center}
    \leavevmode
    \epsfxsize=7.8cm
    \epsfbox[75 75 580 1170]{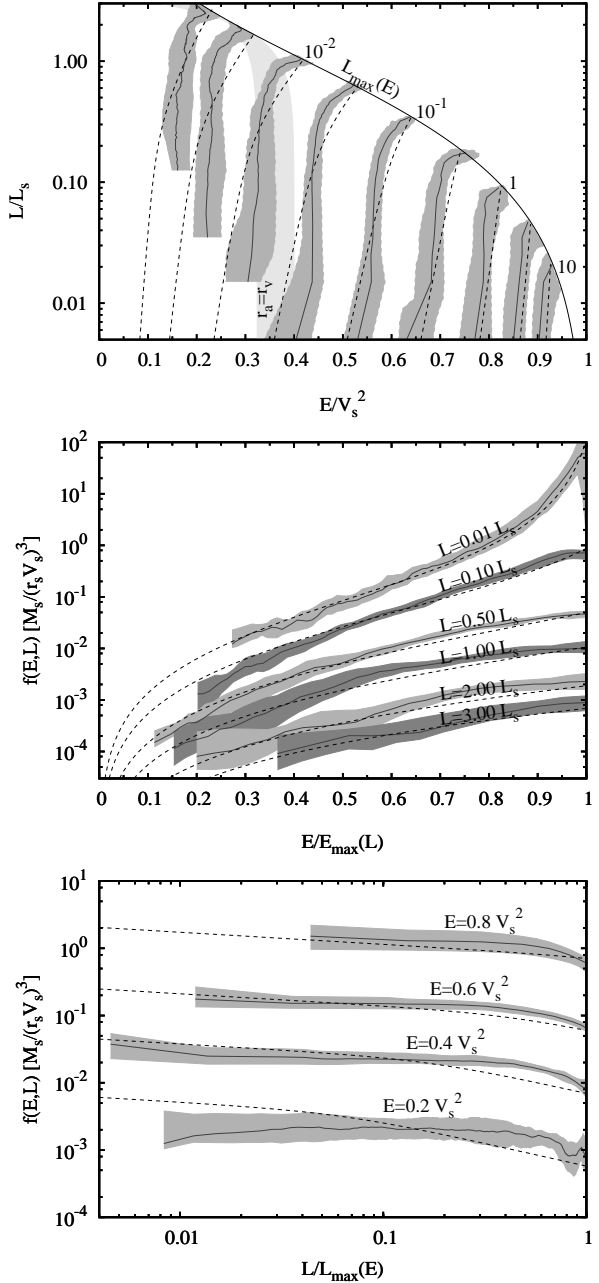}
\end{center}
\caption{Comparison between the DF measured for DM particles
inside the virial sphere of the simulated haloes and the model with parameters adjusted to
the median anisotropy, $\beta_{0}=0.09$, $\beta_{\infty}=0.34$ and
$L_{0}=0.198\,L_{\rm s}$. Solid lines and gray areas stand for median profiles and interquartile
ranges of the DF measured in the halo sample, whereas the dashed lines correspond to
the model. The light gray area in the background of the upper diagram indicates
the points of vanishing radial velocity at the virial radius $r_{\rm v}$. Its boundaries
are fixed by the first and third quartile of virial radii in the halo sample.}
\label{DF_comp}
\end{figure}

Once the model parameters are adjusted the DF
can be compared with its counterpart measured from the simulation.
Fig.~\ref{DF_comp} shows this comparison in terms of a contour map
and the profiles for constant angular momentum or energy. Dark gray
regions in all panels indicate the interquartile range of the DF values
within the halo sample. The lighter gray area in the background of the upper
diagram marks the points of vanishing radial velocity at the virial radius
$r_{\rm v}$. Its boundaries are fixed by the first and third quartile of virial
radii in the halo sample, $4.1r_{\rm s}$ and $6.0r_{\rm s}$ respectively.

Although some deviations of the model (dashed lines) from the results of
the simulations are visible, in general the theoretical profiles are
included within the interquartile range or lie very close to its boundaries.
As expected, the strongest discrepancy between the model and the
simulation is present in the part of the energy-angular momentum plane
populated by the particles with orbits extending beyond the virial sphere
(the area to the left of the $r_{\rm a}=r_{\rm v}$ line). However,
given that this is the only part of the energy-angular momentum
plane affected by the infalling material, we think that the observed
differences are acceptable.

\begin{figure}
\begin{center}
    \leavevmode
    \epsfxsize=8cm
    \epsfbox[75 75 590 612]{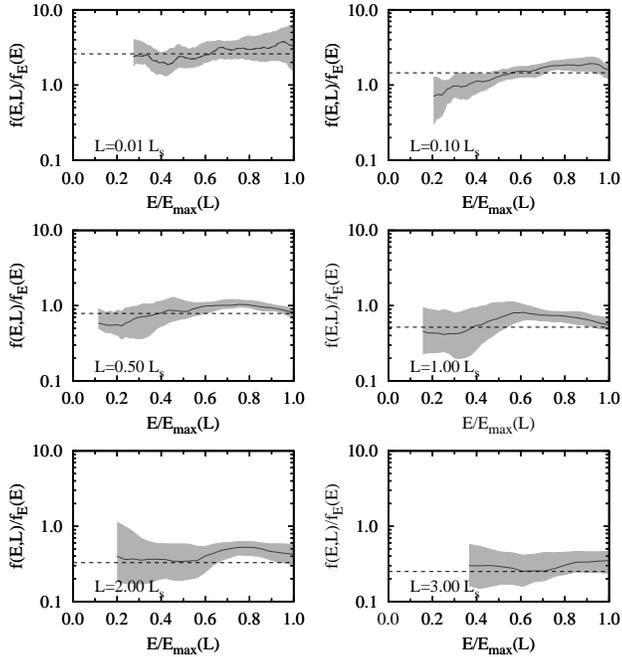}
\end{center}
\caption{The ratio of the DF inferred from the 36 simulated haloes to
the energy part of the DF model with parameters adjusted to the median
anisotropy profile. Each panel shows the profile for the constant value of angular momentum
indicated in the lower left corner. Solid lines and gray areas represent the median profiles
and interquartile ranges respectively. The dashed lines indicate the values
of $f_{L}(L)$ given by the ansatz (\ref{f_L}).}
\label{separability}
\end{figure}

\subsection{The separability of the distribution function}
A critical point of the derivation of the DF presented
in the previous section was the factorization introduced by equation (\ref{DF_mod_gen}).
In order to inspect the robustness of this assumption we propose a simple test.
We calculate the ratio of the DF from the simulation
to the energy part of the DF model with parameters
adjusted to the anisotropy profile from the simulation. Under the assumption
that the real DF is factorizable in energy and angular
momentum, we can expect that the resulting ratio should be a weak
function of energy equal to $f_{L}(L)$ given by (\ref{f_L}).
Fig.~\ref{separability} shows that the variations of this ratio with respect to $f_{L}(L)$
are of the same order as the width of the interquartile range which means that
separability is acceptable from the statistical point of view.
A small systematic deviation can be seen for $L\sim 0.1\,L_{\rm s}$.
However, this is certainly a local feature since this trend is not
repeated in other profiles. Let us emphasize that this test of separability
depends strongly on the reliability of $f_{L}(L)$. One can imagine that an incorrect
form of $f_{L}(L)$ would likely lead to a negative result of the test,
whether $f(E,L)$ is separable or not. On the contrary, a positive result
of such a test in our case means that not only is the assumption of factorization
valid but the approximation for $f_{L}(L)$ is reasonable as well.

\subsection{Velocity moments}

\begin{figure}
\begin{center}
    \leavevmode
    \epsfxsize=8cm
    \epsfbox[75 75 590 612]{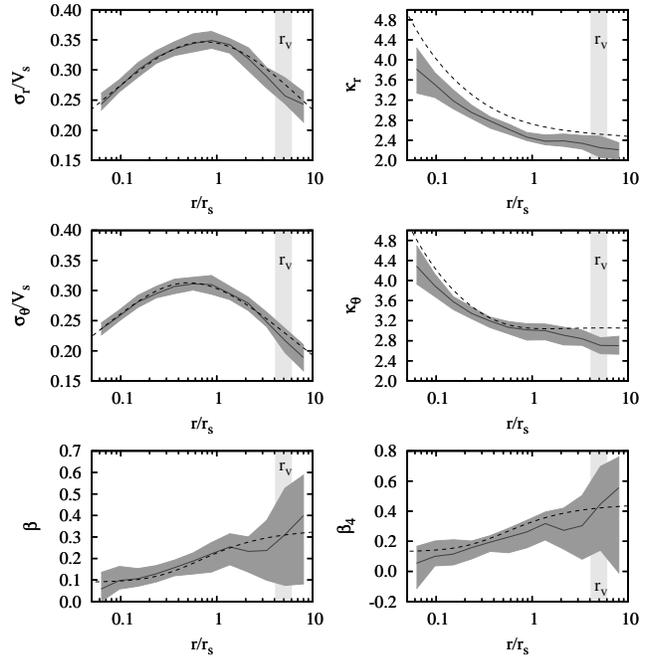}
\end{center}
\caption{Velocity moments of the radial and tangential velocity (top and middle panels) and
the anisotropy of the dispersion tensor $\beta(r)$ and the fourth velocity moment tensor
$\beta_{4}(r)$ defined by (\ref{beta_4}). The solid line and the dark gray area are the median 
and the interquartile range of the profiles obtained for individual haloes and rescaled
by $r_{\rm s}$ and $V_{\rm s}$ inferred from fitting the NFW profile. Dashed lines are
the profiles inferred from
the DF model with parameters adjusted to the median anisotropy measured from the simulation. In the
case of anisotropy $\beta(r)$ (lower left panel) this line shows the fit of the model.}
\label{moments}
\end{figure}

Further comparison between the simulation and the DF model
can be done in terms of velocity moments. This is depicted in
Fig.~\ref{moments} where the dispersion and kurtosis of the radial and tangential
velocity are plotted. In the bottom part of this figure we show the
profiles of the anisotropy $\beta(r)$ and $\beta_{4}$ parameter which measures
the anisotropy of a tensor of the fourth velocity moment. By analogy with the
parameter $\beta(r)$ we defined $\beta_{4}(r)$ in the following way
\begin{equation}\label{beta_4}
	\beta_{4}(r)=1-\frac{\langle v_{\theta}^{4}\rangle(r)}{\langle v_{r}^{4}\rangle(r)}.
\end{equation}
The dashed lines in each panel of Fig.~\ref{moments} are the model predictions,
except for the $\beta(r)$ profile (lower left panel) which is a fit of the model
providing constraints
on parameter values given in the previous subsection. Theoretical dispersion profiles
coincide very well with the profiles from the simulation. We notice quite
a good agreement also
for the $\beta_{4}(r)$ parameter. On the other hand, theoretical profiles of
the kurtosis are systematically biased towards higher values, but typically by less
than 10 percent. Nevertheless, their shapes clearly recover the shapes of the
median profiles from the simulation. Moreover, for both the
radial and tangential velocity
a characteristic growth of $\kappa$ from value $\lesssim 3$ around the virial
radius up to $\gtrsim 4$ in the halo centre, is seen.

Although the kurtosis bias is enclosed within acceptable
limits (kurtosis is known to be sensitive to any noise), it would be desirable
to find out the reason for this behaviour. Since our statistical samples consist of
$10^{4}-10^{5}$ particles per radial bin, we ruled out a possibility
of a bias of the kurtosis estimator (see {\L}okas \& Mamon 2003).
We also excluded the possibility that this is caused by some specific
assumptions of the model. For example, changing the NFW density distribution
to the 3D Sersic profile, which fits the simulation data even better
(Navarro et al. 2004; Merritt et al. 2005; Prada et al. 2006),
we still encounter the same bias. In addition,
perturbing the model parameters of $f_{L}(L)$ does not explain the
situation either.

We therefore conclude that the slight discrepancy in the predictions
of our model concerning the kurtosis
must signify reaching the limitations of the theoretical approach
based on using the global, smooth
gravitational potential of a system. We suppose that the problem is caused
by the presence of substructures which perturb locally the trajectories
of particles with respect to the orbits determined by the
global potential of a halo. What one gets from the simulation is really
a convolution of the velocity distribution expected from the model involving
a global potential with the distribution of velocity perturbations occurring
due to density fluctuations. The estimation of the importance of this effect is
a complicated task since the perturbation of the particle orbit depends on
many variables, such as the distribution of substructures, softening
of the potential and particle velocity. However, some qualitative
conclusions can be drawn. First, note that low-velocity particles
are affected by the density perturbations more strongly. Consequently,
the peak of the resulting velocity distribution is suppressed and the
tails are preserved which may effectively decrease the kurtosis
(see Fig.~\ref{moments}). Second, the effect of the perturbation on the
velocity dispersion is a higher order correction compared to the dispersion
obtained for a system with a global potential. This means that the resulting
dispersion
profiles are barely changed and they are still expected to coincide well
with theoretical predictions.

\begin{figure*}
\begin{center}
    \leavevmode
    \epsfxsize=17cm
    \epsfbox[50 50 1140 1170]{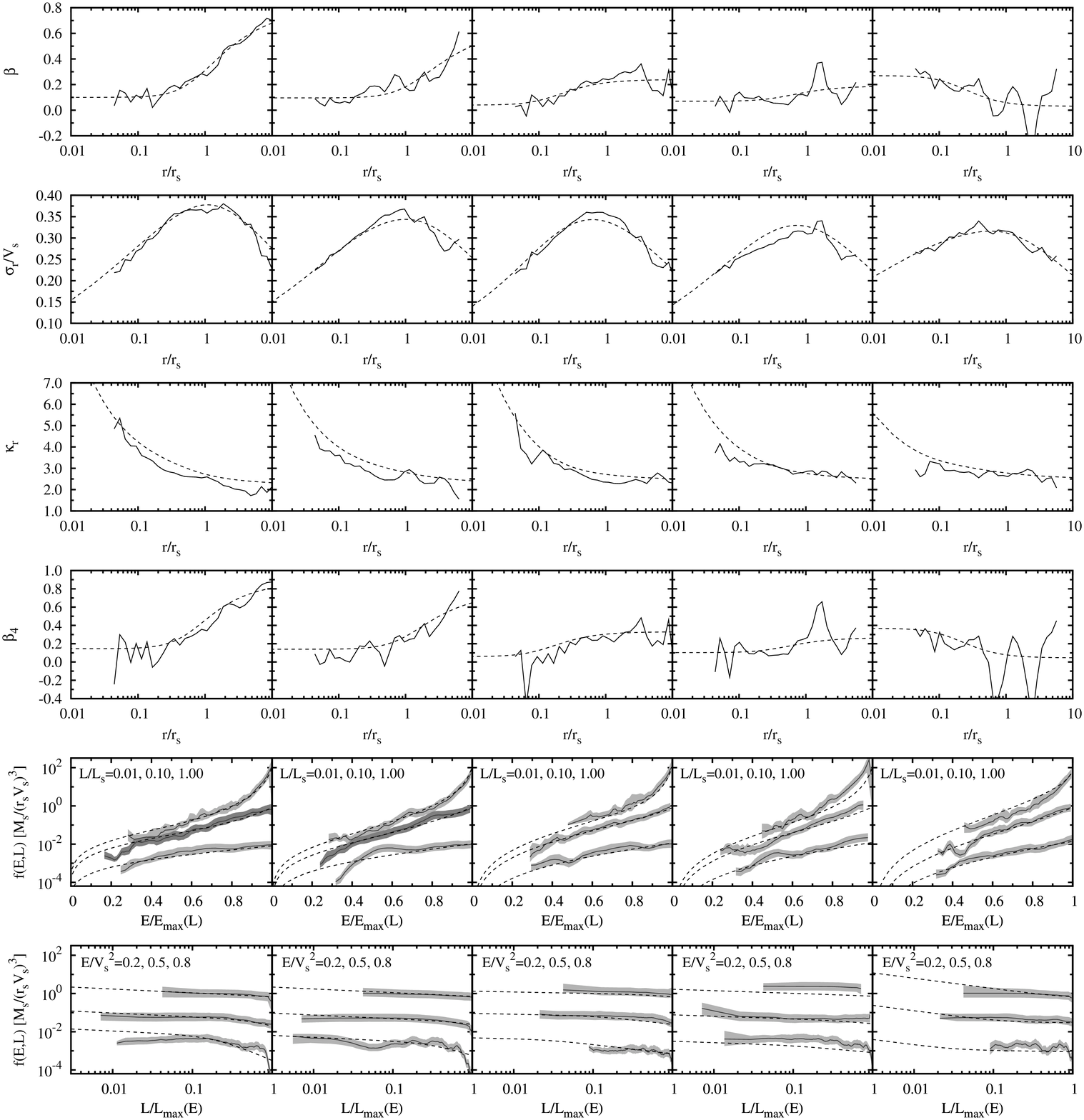}
\end{center}
\caption{Comparison between the model of the DF and the
simulation for 5 haloes (in columns) in terms of the anisotropies
$\beta(r)$ and $\beta_{4}(r)$, the dispersion and kurtosis of the radial velocity
and the profiles of the DF for different values of angular momentum or energy.
Solid and dashed lines show respectively the results from the simulation and
the predictions of the model with parameters adjusted to the anisotropy profile.
Gray areas and lines in the panels of two bottom rows indicate the interquartile
ranges and the medians of the DF for three fixed values of angular momentum or energy
given in the upper left corner of each panel (with lower profiles corresponding
to higher angular momentum or lower energy).}
\label{4haloes}
\end{figure*}

It seems an intriguing issue that the profile of tangential kurtosis
signifies Gaussianity of the velocity distribution at radii around $r_{\rm s}$ where the
logarithmic slope of the density profile is equal to $-2$. One could suppose that some
signatures of the so-called isothermal sphere are locally present. Interestingly,
this statement is also supported by the shape of the DF for
$E\leq \Psi(r_{s})\approx 0.7\,V_{s}^{2}$ that is the energy range of particles at $r_{s}$.
Referring to the middle panel of Fig.~\ref{DF_comp} it is easy to notice that
the DF grows exponentially with energy, as expected for systems not very different
from the isothermal sphere.
The distribution of the radial velocity, on the other hand, takes the Gaussian
form for radii around $0.3 \, r_{\rm s}$. This difference could be a consequence
of the non-vanishing anisotropy parameter, which is not accounted for in the
classical formulation of the isothermal model (Binney 1982; Binney \& Tremaine 1987).

So far we tested the DF model for a typical halo
associated with the median properties of our halo sample. In order to check
the applicability of our model more extensively we repeat such comparison for
single haloes. The DF in this case is expected to differ from
one halo to another due to the observed variety of anisotropy profiles. Results of this
analysis are summarized in Fig.~\ref{4haloes}. To save space
we included only five haloes with representative, rather different
anisotropy profiles (upper panels), from the most strongly increasing profile in
the left panel to a decreasing one on the right-hand side. The second and fifth
panels correspond to the haloes for which contour maps of the DF are shown in
Fig.~\ref{halo_1_DF} (the top and bottom panel respectively). We restricted
the number of profiles to those most essential: we plot the
dispersion and kurtosis of the radial velocity and the anisotropies $\beta(r)$ and $\beta_{4}(r)$.
We also show the profiles of the DF for three values of angular momentum
or energy. In all panels the solid lines represent simulation results, whereas
dashed lines are the predictions of the model. As in Fig.~\ref{moments}, dashed
lines in the case of parameter $\beta(r)$ indicate best fitting profiles of the model.
Gray regions in the panels of two bottom rows mark the interquartile ranges of the
DF which describe the scatter of points resulting from the FiEstAS
algorithm.

From the analysis of Fig.~\ref{4haloes} we conclude that all profiles, regardless
of the anisotropy, are very well reproduced by our model of the DF.
In general, the theoretical DF does not exceed the limits of
the interquartile range or lies very close to its boundaries (see two bottom rows of
panels in Fig.~\ref{4haloes}). Surprisingly,
we find that the agreement is usually almost equally good when the model
is applied to the haloes with massive substructures which were rejected from our
sample. This is certainly good news for the future applications of our DF to the
dynamical modelling of galaxy clusters
which very often display signatures of recent major mergers.

\subsection{Analytical approximation of the distribution function}

\begin{figure}
\begin{center}
    \leavevmode
    \epsfxsize=8cm
    \epsfbox[75 75 580 420]{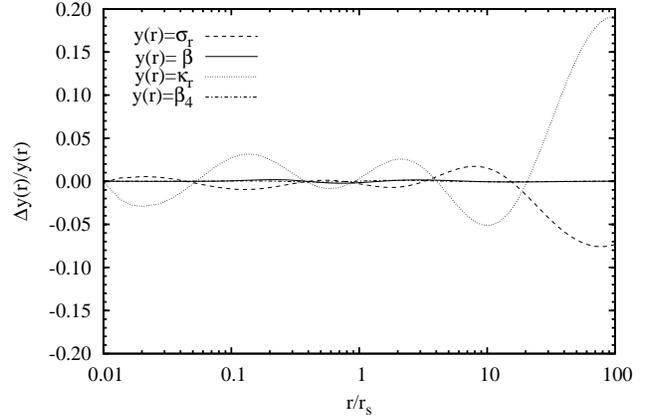}
\end{center}
\caption{Relative errors of velocity moments and anisotropies inferred from
the DF obtained with the analytical approximation of $f_{E}(E)$ given
by (\ref{f_E_appr}). All profiles were compared with the results of exact calculations
summarized in Appendix C.}
\label{errors}
\end{figure}

\begin{table}
\begin{center}
\begin{tabular}{cll}
            & $\beta_{0}=0.09$       &  $\beta_{0}=0$       \\
parameter   & $\beta_{\infty}=0.34$  &  $\beta_{\infty}=0.5$  \\
\hline
$C$                   &      $0.000738$  & $0.001663$ \\
$E_{0}$               &      $0.12903$   & $0.14124$ \\
$E_{1}$               &      $0.078$     & $0.07$ \\
$\alpha_{1}$          &      $2.10 $      & $1.74$ \\
$E_{2}$               &      $0.071$     & $0.085$ \\
$\alpha_{2}$          &      $2.47$      & $3.01$ \\
\hline
\end{tabular}
\end{center}
\caption{Values of the parameters used in the approximation
of the energy part of the DF (\ref{f_E_appr}).
The first column lists the parameters. The second column gives the
parameter values for the model fitted to the anisotropy for the average halo from
the simulation ($\beta_{0}=0.09$ and $\beta_{\infty}=0.34$) and the
third one gives the values which reproduce the DF for the anisotropy profile
(\ref{ML_beta}) ($\beta_{0}=0$ and $\beta_{\infty}=0.5$). In both
cases $L_{0}=0.198\,L_{\rm s}$ was assumed.}
\label{parameters}
\end{table}

The DF discussed in the first subsection is typical in a sense that
it describes the statistical macrostate of DM particles
in a typical massive halo. With future applications in mind we decided to provide an
analytical approximation for the energy part $f_{E}(E)$ of this DF which could be used as a substitute
for a rather complicated procedure described in Appendix B. We found
that the following expression reproduces the numerical DF with good accuracy
\begin{eqnarray}
	f_{E}(E) & = & C\exp \left( {E\over E_{0}} \right )E^{\alpha_{1}/[1+\exp(E/E_{1})]}\nonumber \\
	& & \times(1-E)^{-\alpha_{2}/\{1+\exp[(1-E)/E_{2})]\}} \label{f_E_appr},
\end{eqnarray}
where the values of the parameters are listed in the middle column
of Table~\ref{parameters}. For completeness we recall that the angular momentum part of the DF is given by
(\ref{f_L}) with $\beta_{0}=0.09$, $\beta_{\infty}=0.34$ and $L_{0}=0.198\,L_{\rm s}$. We
have verified that the errors of the dispersion, kurtosis and both anisotropies, when
using this approximate formula in the integrals for velocity moments, do not exceed
5 percent within the radial range $(0.01r_{\rm s},30r_{\rm s})$ (see
Fig.~\ref{errors}). Note that the general form of expression (\ref{f_E_appr})
can be effectively used to approximate the DF model also for other sets of
parameters. As a second example we include in the third column of the
Table also the parameters of a model with $\beta_{0}=0$,
$\beta_{\infty}=0.5$ and $L_{0}=0.198\,L_{\rm s}$ which mimics the anisotropy profile
(\ref{ML_beta}) with $r_{1/4}=0.9\,r_{\rm s}$.

\section{Discussion}

We have studied the DF of DM particles inside the virial
spheres of the haloes of mass $10^{14}$--$10^{15}\Msun$ formed in the standard
$\Lambda$CDM cosmological $N$-body simulation.
In the first part of the paper we presented results of the calculation of the DF from
the simulation in the form most suitable for comparison with
theoretical models. Then we proposed a phenomenological model of the DF.
The model in its part dependent on angular momentum involves three free parameters
which specify the anisotropy
profile, namely its asymptotic values and the scale
of transition between them. We demonstrated that this parametrization
is sufficient to reproduce accurately the simulation results in terms of
velocity moments as well as the DF itself. The only discrepant point we encountered
was a small but statistically significant bias of the theoretical kurtosis with respect to its
profiles measured from the simulation. This is probably caused by the
presence of substructures perturbing the trajectories of low-velocity particles.

In section 5 we showed that the velocity distribution of a typical halo changes from
a flat-topped distribution ($\kappa<3$) in the outer part to a peaked one
($\kappa>3$) near the centre. This behaviour was noticed and discussed by others before
(e.g. Kazantzidis, Magorrian \& Moore 2004; Wojtak et al. 2005).
The analysis of the DF presented here suggests that
this property of the velocity distribution is correlated with the
profile of the anisotropy increasing with $r$: $\beta(r)$ profiles growing faster with $r$
imply more rapid growth of the kurtosis towards the centre.

\begin{figure}
\begin{center}
    \leavevmode
    \epsfxsize=8cm
    \epsfbox[75 75 590 350]{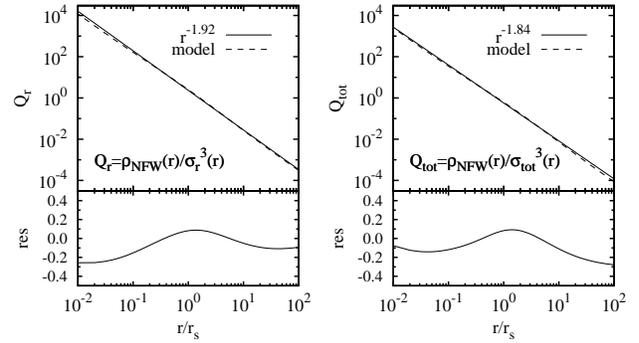}
\end{center}
\caption{Radial profiles of $\rho(r)/\sigma_{r}(r)^{3}$ (left panel) and
$\rho(r)/\sigma_{\rm tot}(r)^{3}$ (right panel). The dashed lines show predictions of the
model DF with the NFW density profile and parameters
fitted to the median anisotropy profile shown in Fig.~\ref{beta_sing}. The solid lines
plot power-law functions with logarithmic slopes from Dehnen \&
McLaughlin (2005). In the lower panels we show relative residuals.}
\label{Q_profile}
\end{figure}

As demonstrated by Taylor \& Navarro (2001), the profile of the phase-space density
$Q(r)=\rho(r)/\sigma(r)^{3}$
in DM haloes is well fitted by a power-law function. It seems that
the status of this relation is as well established as the NFW fit of the
density profile. We checked that $Q(r)$ profiles inferred from
the DF model with parameters adjusted to the median anisotropy (see section 5)
coincide well with the corresponding power-law functions (see
Fig.~\ref{Q_profile}). In this comparison we assumed the logarithmic
slopes of $-1.92$ (in the case of the dispersion of the radial velocity) and $-1.84$
(in the case of the total dispersion), the values obtained from the
simulations by Dehnen \& McLaughlin (2005).
The relative residuals of both $Q(r)$ profiles are of the same order as the
scatter of points from the simulations in fig.~1 of Dehnen \& McLaughlin
(2005). Note that this happens when the DF model is tuned to the mean
trend of the $\beta(r)$ parameter. Therefore one could suspect that both relations,
the mean profile of the anisotropy and $Q(r)\propto r^{-\gamma}$, are two aspects
of some deeper relation. A more general parametrization of the DF might provide some
insights towards a more fundamental understanding of this phenomenon.

%

Although the whole analysis presented in this paper was done in the framework of the NFW
density profile, the equations for the numerical inversion (\ref{inv_final}) were derived
for an arbitrary density distribution. Using this general form one can
immediately obtain a family of DFs with our general anisotropy profiles for
any potential-density pair. For the commonly used density profiles it is
easy to introduce the phase space units analogous to $r_{\rm s}$,
$V_{\rm s}$ and $M_{\rm s}$ in our case. This reduces the role of the parameters of the
density profile
to scaling properties so that the final DF model would not explicitly
depend on them.

Given our very general parametrization of the $\beta(r)$ profile,
our DF model is expected to provide some impact
on the solution of the classical problem of mass-anisotropy degeneracy for
spherical systems. In order to obtain more reliable estimates of mass profiles,
one could assume the anisotropy profile from the simulation and keep the density
profile as the only degree of freedom of the DF. One could then apply
the maximum likelihood approach of projected DF, as described e.g.
by Mahdavi \& Geller (2004). A more advanced and simulation-independent approach
would be to treat the anisotropy profile as an unknown quantity,
described by the three parameters introduced in our formulation.
As a result one would obtain an estimate of the mass profile as well as
anisotropy. Both methods require an additional study of the DF in
projection and extensive tests on mock data sets. This will be the subject of
our follow-up papers.

\section*{Acknowledgments}

The simulations have been performed at the Altix of the LRZ Garching.
RW and E{\L} are grateful for the hospitality of Astrophysikalisches Institut
Potsdam and Institut d'Astrophysique de Paris where part of this work was done.
RW thanks A. Knebe for helpful advice and G. Bou\'e for fruitful discussions.
This work was partially supported by the Polish Ministry of Science and Higher Education
under grant NN203025333 as well as by the Polish-German exchange
program of Deutsche Forschungsgemeinschaft and the Polish-French collaboration
program of LEA Astro-PF.

\appendix
\onecolumn

\section{The volume of the hypersurface}
The volume $g(E,L)$ of the hypersurface $S_{E\,L}({\mathbf v},{\mathbf r})$
of constant energy and angular momentum is defined by the integral
\begin{equation}\label{g_der1}
	g(E,L)\textrm{d}E\textrm{d}L=
	\int_{S_{E\,L}({\mathbf v},{\mathbf r})\textrm{d}E\textrm{d}L}
	\!\!\!\!\textrm{d}^{3}v \: \textrm{d}^{3}r.
\end{equation}
Introducing spherical coordinates and changing variables into $E$, $L$
and radius $r$ one gets
\begin{equation}\label{g_der12}
	g(E,L)\textrm{d}E\textrm{d}L=8\pi^{2}L
	\textrm{d}E\textrm{d}L\oint\:\frac{\textrm{d}r}
	{|v_{r}(E,L,r)|},
\end{equation}
where $v_{r}$ is the radial velocity and the integral is equal to the radial
period of the orbit. Using the radii of the pericentre $r_{\rm p}$ and the apocentre
$r_{\rm a}$, one can rewrite the final formula for $g(E,L)$ in the following
way
\begin{equation}\label{g_der2}
	g(E,L)=16\pi^{2}L\int_{r_{\rm p}}^{r_{\rm a}}
	\frac{\textrm{d}r}{\sqrt{2\Psi(r)-2E-L^{2}/r^{2}}}.
\end{equation}
Integrating $g(E,L)$ over the angular momentum we get the volume $g_{E}(E)$
of the hypersurface of constant energy
\begin{equation}\label{gE_der1}
	g_{E}(E)=16\pi^{2}\int_{0}^{L_{\rm max}(E)}L\textrm{d}L
	\int_{r_{\rm p}(L)}^{r_{\rm a}(L)}
	\frac{\textrm{d}r}{\sqrt{2\Psi(r)-2E-L^{2}/r^{2}}}.
\end{equation}
Changing the order of the integrals and performing the integral over
the angular momentum one obtains
\begin{equation}\label{gE_der2}
	g_{E}(E)=16\pi^{2}\int_{0}^{r_{\rm max}(E)}
	\sqrt{2(\Psi(r)-E)}r^{2}\textrm{d}r,
\end{equation}
where $r_{\rm max}(E)$ is the apocentre radius of the radial orbit
($E=\Psi(r_{\rm max})$).

Considering a system of finite size $V({\mathbf v},{\mathbf r})$
in phase space, one has to recalculate $g(E,L)$ with the realistic
hypersurface of constant $E$ and $L$ given by the Cartesian
product $S_{E\,L}({\mathbf v},{\mathbf r})\times V({\mathbf v},{\mathbf r})$.
In particular, for a spherical system with a boundary in the form of a
sphere of radius $r_{\rm v}$ the upper limit of the integral in
(\ref{g_der2}) and (\ref{gE_der2}) must be replaced by
$\textrm{min}\{r_{\rm a},r_{\rm v}\}$ and $\textrm{min}\{r_{\rm a},r_{\rm max}(E)\}$
respectively.

\section{The energy part of the distribution function}
The energy part of the DF $f_{E}(E)$ introduced in section 4 is
related to the density profile by
\begin{equation}\label{inv_1}
	\rho(r)=\int\!\!\!\int\!\!\!\int f_{E}(E)
	\Big(1+\frac{L^{2}}{2L_{0}^{2}}\Big)^{-\beta_{\infty}+\beta_{0}}
	L^{-2\beta_{0}}\textrm{d}^{3}v.
\end{equation}
Although the main part of this paper concerns the DF consistent
with the NFW profile, we keep within the appendix a general density $\rho(r)$ so that
the final formulae of inversion could be applied to any potential-density pair.

Changing variables in the integral (\ref{inv_1}) into the energy and angular momentum
one gets
\begin{equation}\label{inv_2}
	\rho(r) = 2^{3/2-\beta_{0}}\pi r^{-1}L_{0}^{1-2\beta_{0}}
	\int_{0}^{\Psi}f_{E}(E)\textrm{d}E
	\int_{0}^{x}\frac{(1+\lambda)^{-\beta_{\infty}+\beta_{0}}\lambda^{-\beta_{0}}}
	{\sqrt{x-\lambda}}\textrm{d}\lambda,
\end{equation}
where $x=r^{2}(\Psi-E)/L_{0}^{2}$ and $\lambda=L^{2}/(2L_{0}^{2})$. The integral over
the $\lambda$ variable is evaluated analytically so that (\ref{inv_2}) can be
rewritten in the form
\begin{equation}\label{inv_3}
	\rho(r)r^{2\beta_{0}} = (2\pi)^{3/2}2^{-\beta_{0}}
	\frac{\Gamma(1-\beta_{0})}{\Gamma(3/2-\beta_{0})}
	\int_{0}^{\Psi}f_{E}(E)(\Psi-E)^{1/2-\beta_{0}}K(\Psi,E)\textrm{d}E
\end{equation}
with a kernel of the integral given by
\begin{equation}\label{kernel}
	K(\Psi,E)=(1+x)^{-\beta_{\infty}+\beta_{0}}
	{}_{2}F_{1}(1/2,\beta_{\infty}-\beta_{0},3/2-\beta_{0},x/(1+x)),
\end{equation}
where ${}_{2}F_{1}$ stands for the hypergeometric function. Equation
(\ref{inv_3}) is a Volterra integral of the first kind. In the general case
of models with varying anisotropy, when $\beta_{\infty}\neq\beta_{0}$
and $L_{0}<\infty$, it has no analytical solution for $f_{E}(E)$ due
to the complexity of expression (\ref{kernel}). However, as shown by Cuddeford
\& Louis (1995), this kind of integral can be quite easily inverted
numerically. Below we adapt their method to our problem.

For $E\rightarrow 0$ and $\Psi\gg E$ the integral kernel can encounter a singularity,
i.e. $K(\Psi,E)\propto E^{\beta_{\infty}-\beta_{0}}$. In order to avoid this
behaviour, we define a smooth integral kernel $\widehat{K}(\Psi,E)$ which is free
of such a feature
\begin{equation}\label{kernel_sm}
	\widehat{K}(\Psi,E)=E^{-\beta_{\infty}+\beta_{0}}K(\Psi,E).
\end{equation}
By analogy we introduce a smooth energy part of the DF
which is a regular function for energy approaching $0$
\begin{equation}\label{f_E_sm}
	\widehat{f}_{E}(E)=E^{3/2-\nu}f_{E}(E).
\end{equation}
The value of $\nu$ is determined from the limit of $\rho(r)$
at small potential ($\Psi\ll 1$ and
$r\rightarrow \infty$) in equation (\ref{inv_3}), for which
\begin{equation}
	\rho(r)r^{2\beta_{0}} \propto
	\Psi(r)^{\nu-\beta_{0}}r^{2(-\beta_{\infty}+\beta_{0})},
\end{equation}
which immediately gives $\nu$ in terms of $\beta_{0}$, $\beta_{\infty}$ and
the asymptotic slopes of the potential and density. In particular, for the
NFW density profile one gets $\nu=3-2\beta_{\infty}+\beta_{0}$.

Using formulae (\ref{kernel_sm}) and (\ref{f_E_sm}) we can rewrite equation
(\ref{inv_3}) in the following form
\begin{equation}\label{inv_4}
	\widehat{\rho}(r)=C_{\beta_{0}}\!\!\int_{0}^{\Psi}\!\!\widehat{f}_{E}(E)
	\widehat{K}(\Psi,E)(\Psi-E)^{1/2-\beta_{0}}
	E^{\nu-3/2+\beta_{\infty}-\beta_{0}}\textrm{d}E,
\end{equation}
where $\widehat{\rho}=\rho(r)r^{2\beta_{0}}$ and $C_{\beta_{0}}$ stands for all
coefficients in front of the integral (\ref{inv_3}). Following
Cuddeford \& Louis (1995) we introduce discrete vectors of the potential
$\Psi_{j}=j\epsilon$, radius $r_{j}=r(\Psi_{j})$ and density
$\widehat{\rho}_{j}=\rho(r_{j})r_{j}^{2\beta_{0}}$, where $j$ is an integer
number and $\epsilon=1/j\ll 1$. For any $\widehat{\rho}_{j}$ we can split the integral
(\ref{inv_4}) into a sum
\begin{equation}\label{inv_5}
	\widehat{\rho}_{j} = C_{\beta_{0}}\sum_{i=1}^{i=j}
	\int_{(i-1)\epsilon}^{i\epsilon}\widehat{f}_{E}(E)
	\widehat{K}(j\epsilon,E)(j\epsilon-E)^{1/2-\beta_{0}}
	E^{\nu-3/2+\beta_{\infty}-\beta_{0}}\textrm{d}E.
\end{equation}
In order to apply any numerical algorithm to invert equation (\ref{inv_5})
with respect to $\widehat{f}_{E}(E)$, one has to assume that $\epsilon$ is
sufficiently small so that the variations of $\widehat{f}_{E}(E)$ and
$\widehat{K}(j\epsilon,E)$ within subsequent integration ranges are negligible.
Then one can approximate both functions by their values at $(i-1/2)\epsilon$,
i.e. the middle points of the energy range. This approach was used by
Cuddeford \& Louis (1995) and favoured over other methods involving
higher order interpolation (see e.g. Saha 1992). Applying this approximation
to equation (\ref{inv_5}) we get
\begin{equation}\label{inv_6}
	\widehat{\rho}_{j}=C_{\beta_{0}}\sum_{i=1}^{i=j}\widehat{f}_{E\,i}
	\widehat{K}(j\epsilon,(i-1/2)\epsilon)I_{ij}
\end{equation}
with $\widehat{f}_{E\,i}=\widehat{f}_{E}((i-1/2)\epsilon)$ and the matrix $I_{ij}$
defined in the following way
\begin{eqnarray}\label{I_ij}
	I_{ij} & = & \int_{(i-1)\epsilon}^{i\epsilon}
	E^{\nu-3/2+\beta_{\infty}-\beta_{0}}(j\epsilon-E)^{1/2-\beta_{0}}
	\textrm{d}E\nonumber \\
	I_{ij} & = & (j\epsilon)^{\nu+\beta_{\infty}-2\beta_{0}}
	[B_{i/j}(\nu+\beta_{\infty}-\beta_{0}-1/2,3/2-\beta_{0})
	-B_{(i-1)/j}(\nu+\beta_{\infty}-\beta_{0}-1/2,3/2-\beta_{0})],
\end{eqnarray}
where $B_{z}(x,y)$ is the incomplete beta function. As shown by Cuddeford
\& Louis (1995), the solution of (\ref{inv_6}) for $\widehat{f}_{E\,i}$ can be
obtained by evaluating iteratively the following expression
\begin{equation}\label{inv_final}
	\widehat{f}_{E\,j}=\frac{\widehat{\rho}_{j}/C_{\beta_{0}}-\sum_{i=1}^{i=j-1}
	\widehat{K}(j\epsilon,(i-1/2)\epsilon)I_{ij}\widehat{f}_{E\,i}}
	{\widehat{K}(j\epsilon,(j-1/2)\epsilon)I_{jj}}
	\end{equation}
with the initial value $\widehat{f}_{E\,1}$ given by
\begin{equation}\label{inv_final_init}
	\widehat{f}_{E\,1}=\frac{\widehat{\rho}_{1}/C_{\beta_{0}}}
	{\widehat{K}(\epsilon,\epsilon/2)I_{11}}.
\end{equation}

\section{Velocity moments}
All non-vanishing moments of the radial or tangential velocity at a given
radius $r$ are defined by the integral
\begin{equation}\label{vel_mom_1}
	\langle v_{i}^{2n}\rangle(r) =
	\frac{1}{\rho(r)} \int\!\!\!\int\!\!\!\int f_{E}(E)
	\Big(1+\frac{L^{2}}{2L_{0}^{2}}\Big)^{-\beta_{\infty}+\beta_{0}}
	L^{-2\beta_{0}}v_{i}^{2n}\textrm{d}^{3}v,
\end{equation}
where $i$ is a subscript indicating the velocity component in spherical
coordinates and $n$ is an integer number. By analogy with steps
(\ref{inv_2}) and (\ref{inv_3}) from Appendix B, one can
evaluate the angular momentum part of the integral and simplify the expression
to the following form
\begin{equation}\label{vmom}
	\langle v_{i}^{2n}\rangle(r) = \frac{
	2^{3/2+n-\beta_{0}}\pi}
	{r^{2\beta_{0}}\rho(r)}
	\int_{0}^{\Psi}f_{E}(E)(\Psi-E)^{1/2+n-\beta_{0}}K_{i}(\Psi,E)\textrm{d}E,
\end{equation}
where the kernel functions for the radial ($i=r$) and tangential
($i=\theta=\phi$) component of the velocity field are given
respectively by
\begin{eqnarray}
	K_{r}(\Psi,E) & = &
	\frac{\Gamma(3/2+n)\Gamma(1-\beta_{0})}{(1/2+n)\Gamma(3/2+n-\beta_{0})}
	(1+x)^{-\beta_{\infty}+\beta_{0}}
	{}_{2}F_{1}(1/2+n,\beta_{\infty}-\beta_{0},3/2+n-\beta_{0},x/(1+x))\label{moments_kr} \\
	K_{\theta}(\Psi,E) & = &
	\frac{\Gamma(1/2+n)\Gamma(1+n-\beta_{0})}{\Gamma(1+n)\Gamma(3/2+n-\beta_{0})}
	(1+x)^{-\beta_{\infty}+\beta_{0}}
	{}_{2}F_{1}(1/2,\beta_{\infty}-\beta_{0},3/2+n-\beta_{0},x/(1+x))\label{moments_kt}.
\end{eqnarray}
Once $f_{E}(E)$ is calculated in terms of the $f_{E\,i}$ vector it is easy to evaluate
numerically the integral (\ref{vmom}) and obtain the profile of any velocity moment.

An interesting property of the model is the ratio of any non-vanishing
moment of the tangential velocity to the corresponding moment of the radial velocity
in the limit of small and large radii. Introducing spherical coordinates
in (\ref{vel_mom_1}) and performing the integral with two asymptotes of
$f_{L}(L)$ given by (\ref{f_L_asympt}), one can show that this ratio is
the following function of $\beta_{0}$ or $\beta_{\infty}$
\begin{equation}\label{beta_n_asympt}
	\frac{\langle v_{\theta}^{2n}\rangle}{\langle v_{r}^{2n}\rangle}= \left\{
	\begin{array}{ll}
	\displaystyle{
	\frac{\Gamma(1+n-\beta_{0})}{\Gamma(1-\beta_{0})\Gamma(1+n)} }
	& \textrm{for $r\rightarrow 0$} \\
	 & \\
	\displaystyle{
	\frac{\Gamma(1+n-\beta_{\infty})}{\Gamma(1-\beta_{\infty})\Gamma(1+n)} }
	& \textrm{for $r\rightarrow \infty$.}
	 \\
	\end{array} \right.
\end{equation}

\end{document}